\documentclass[submission]{eptcs}

\providecommand{\event}{ICLP 2025}

\usepackage{framed}
\usepackage{subcaption}

\newlength{\cslhangindent}
{\section*{References}\setlength{\parindent}{0pt}\setlength{\parskip}{5pt}}%
{\par}

\usepackage{tikz}
\usetikzlibrary{shapes.misc,shadows}
\usetikzlibrary{automata, positioning, arrows}
\usepackage{mathpazo}

\newcounter{row}
\newcounter{col}

\newcommand\citep[1]{\cite{#1}}
\newcommand\citet[1]{\cite{#1}}

\usepackage{listings}
\newcommand{\passthrough}[1]{#1}

\begin{document}

\title{%
  Logic Programming with Extensible Types
}

\def\titlerunning{Logic Programming with Extensible Types}
\def\authorrunning{I. Perez \& A. Herranz}

\author{Ivan Perez
\institute{KBR @ NASA Ames Research Center\\
California, USA}
\email{ivan.perezdominguez@nasa.gov}
\and
Angel Herranz
\institute{Universidad Politecnica de Madrid\\
Madrid, Spain}
\email{angel.herranz@upm.es}
}

\maketitle

\begin{abstract}
Logic programming languages present clear advantages in terms of
declarativeness and conciseness. However, the ideas of logic programming
have been met with resistance in other programming communities, and have
not generally been adopted by other paradigms and languages. This paper
proposes a novel way to incorporate logic programming in an existing
codebase in a typed functional programming language. Our approach
integrates with the host language without sacrificing static typing, and
leverages strengths of typed functional programming such as polymorphism
and higher-order. We do so by combining three ideas. First, we use the
extensible types technique to allow values of the host language to
contain logic variables. Second, we implement a unification algorithm
that works for any data structure that supports certain operations.
Third, we introduce a domain-specific language to define and query
predicates. We demonstrate our proposal via a series of examples, and
provide aids to make the notation convenient for users, showing that the
proposed approach is not just technically possible but also practical.
Our ideas have been implemented in the language Haskell with very good
results.

\end{abstract}

\hypertarget{introduction}{%
\section{Introduction}\label{introduction}}

Definitions in imperative and functional languages are structured around
functions and procedures, which are executed by providing the inputs and
evaluating the result. For example, for the Haskell type
\passthrough{\lstinline!data Nat = Zero | Suc Nat!}, we can define
addition as:

\begin{lstlisting}[language=Haskell]
plus :: Nat -> Nat -> Nat
plus Zero    y = y
plus (Suc x) y = Suc (plus x y)
\end{lstlisting}

Although, at an abstract level, functions are relations between sets, it
is not usually possible to treat functions as relations in a language
like Haskell. For example, we cannot use \passthrough{\lstinline!plus!}
to efficiently calculate the subtraction function or to generally
calculate all tuples of inputs and outputs in the
\passthrough{\lstinline!plus!} relation. Contrast this limitation with
how one would write an analogous predicate in a logic programming
language like Prolog:

\begin{lstlisting}[language=Prolog]
plus(zero,   Y, Y).
plus(suc(X), Y, suc(Z)) :- plus(X,Y,Z).
\end{lstlisting}

In Prolog, \passthrough{\lstinline!plus(A,B,C)!} is true if
\passthrough{\lstinline!C!} represents the sum of
\passthrough{\lstinline!A!} and \passthrough{\lstinline!B!}. The
predicate can be used to add numbers, subtract a number from another, or
check if two numbers add up to a given third, or obtain tuples of inputs
and output for which the relation holds: \footnote{We manually stop the
  production of solutions to the last query.}

\begin{lstlisting}[language=Prolog]
?- plus(suc(suc(zero)),B,suc(suc(suc(zero)))).
B = suc(zero).
?- plus(zero,zero,suc(zero)).
false.
?- plus(A,suc(zero),C).
A = zero, C = suc(zero) ;
A = suc(zero), C = suc(suc(zero)) .
\end{lstlisting}

Integrating ideas from logic programming into other languages generally
requires substantial changes to existing codebases and the types used
throughout. Depending on the approach and the engine used to process
queries, it may also lead to loss of static type safety or other
features of the host language (e.g., polymorphism, higher order).

This paper describes a technique for integrating predicates in the style
of logic programming in an existing codebase in a different programming
language. We use Haskell to demonstrate our proposal, but the ideas can
be applied to other languages. We show, with examples, that our approach
requires very little work on the side of the programmer, can capture
many of the use cases of logic programming languages, and can be enabled
by convenient notation. Specifically, the contributions of this paper
are:

\begin{itemize}
\item
  We present an interface for logic programming that facilitates
  introducing logic variables in algebraic datatypes and expressing
  unification constraints (Section
  \ref{logic-programming-with-extensible-types}).
\item
  We show that the proposed approach is applicable to polymorphic types,
  and enables leveraging the host language's mechanisms for type
  inference and higher-order to implement type-safe higher-order logic
  programming (Section \ref{polymorphism-and-higher-order}).
\item
  We extend the language with \emph{cuts}, which allow users to increase
  performance, provide determinism, and encode negation as failure
  (Section \ref{cuts}).
\end{itemize}

Section \ref{implementation} discusses an implementation that
demonstrates our proposal, Section \ref{related-work} details related
work, and Section \ref{future-work} proposes future work.

\hypertarget{background}{%
\section{Background}\label{background}}

\emph{Extensible types} \citep{2023:perez:extensibletypes} are a design
pattern in which a data type is parameterized by a type function that is
applied to every element of the definition. For example, given a type
representing expressions, like
\passthrough{\lstinline!data Expr = Const Double | Add Expr Expr | Neg Expr!},
we define the matching extensible type:

\begin{lstlisting}[language=Haskell]
data ExprF f = ConstF (f Double)
             | AddF (f (ExprF f)) (f (ExprF f))
             | NegF (f (ExprF f))
\end{lstlisting}

If we use the polymorphic type \passthrough{\lstinline!Identity!} as
type function \passthrough{\lstinline!f!}, the resulting representation
is isomorphic to the original \passthrough{\lstinline!Expr!}. Other
parametric types and type functions render different results. For
example, a type that pairs elements with a tuple of
\passthrough{\lstinline!Int!}s can be used to annotate values with the
line and column where they were found in an input file, useful in
compilers to report error information. Applying
\passthrough{\lstinline!Maybe!} or \passthrough{\lstinline!Either!}, an
extensible type makes every element optional, a representation that is
useful in parsing to mark branches of an abstract syntax tree (AST) that
failed to parse. Other type functions enable changing type definitions
to introduce new cases, prune branches, replace elements, etc. The
composition of extensible types can capture language embeddings.

The application of a specific type function to an extensible type does
not determine how it should be interpreted. For example,
\passthrough{\lstinline!Either String!} can be used to annotate failed
AST branches with the reasons why values could not be parsed from an
input file, but also to replace branches in the AST by variables with
the given variable names. This idea will be used in future sections to
replace portions of a datatype with variables in predicate definitions
and logic programming queries.

In the rest of the text, we refer to types that are parameterized in
this manner simply as \emph{extensible types}. Other approaches to
parameterize a type by a type function used in its definition are
further discussed in Section \ref{related-work}.

\hypertarget{logic-programming-with-extensible-types}{%
\section{Logic Programming with Extensible
Types}\label{logic-programming-with-extensible-types}}

This section introduces primitives to define and combine predicates, and
ways to capture relations between values of algebraic datatypes. We
first introduce basic types, and simple primitives and connectives. We
later show how to replace portions of values with logic variables, and
how to express relations involving types with variables.

\hypertarget{goals-primitives-and-boolean-combinators}{%
\subsection{Goals, Primitives and Boolean
Combinators}\label{goals-primitives-and-boolean-combinators}}

The elementary type in our proposal is a \passthrough{\lstinline!Goal!},
which denotes a logic goal or, put simply, something that must be
proven. We keep the type abstract for now and discuss implementation
details later. To interact with \passthrough{\lstinline!Goal!}s, we
provide the function \passthrough{\lstinline!repl :: Goal -> IO ()!}
that, when applied to a \passthrough{\lstinline!Goal!}, produces
possible solutions one by one, similar to the REPL of a logic
programming language. If constraints apply for the goal to hold,
\passthrough{\lstinline!repl!} prints the constraints; otherwise, it
prints ``\passthrough{\lstinline!true.!}'' or
``\passthrough{\lstinline!false.!}''. In this paper, we align queries to
and results from \passthrough{\lstinline!repl!} for readability.

\hypertarget{primitives}{%
\subsubsection{Primitives}\label{primitives}}

We provide \passthrough{\lstinline!succeed :: Goal!}, which holds
without additional constraints, and its counterpart,
\passthrough{\lstinline!fail :: Goal!}, which always fails. We can
evaluate either goal in a session with the
\passthrough{\lstinline!repl!} function, as follows:

\begin{lstlisting}
> repl fail
false.
\end{lstlisting}

\hypertarget{boolean-connectives}{%
\subsubsection{Boolean Connectives}\label{boolean-connectives}}

Our counterparts for the boolean connectives \emph{and} and \emph{or},
which we denote \passthrough{\lstinline!(@@)!} and
\passthrough{\lstinline!(@|)!}, allow users to combine goals:

\begin{lstlisting}[language=Haskell]
(@@) :: Goal -> Goal -> Goal
(@|) :: Goal -> Goal -> Goal
\end{lstlisting}

\hypertarget{example}{%
\paragraph{Example}\label{example}}

The following queries show how we can use the boolean connectives to
combine \passthrough{\lstinline!succeed!} and
\passthrough{\lstinline!fail!}. The results should be straightforward:

\begin{lstlisting}
> repl (succeed @@ fail)
false.
> repl (succeed @@ (fail @| succeed))
true.
\end{lstlisting}

\hypertarget{terms-and-logic-variables}{%
\subsection{Terms and Logic Variables}\label{terms-and-logic-variables}}

To introduce logic variables in values, we apply a type function to
extensible types. We introduce a custom sum type
\passthrough{\lstinline!Term!}, which can represent a logic variable
with a name, or an actual value of a given type:

\begin{lstlisting}[language=Haskell]
data Term a = Var String | Compound a
\end{lstlisting}

When \passthrough{\lstinline!Term!} is applied to an extensible type,
every element inside the latter can potentially be replaced with a logic
variable, allowing us to describe values in which some portions are
concrete and some portions are not.

\hypertarget{example-1}{%
\paragraph{Example}\label{example-1}}

Given the usual encoding of Peano numbers using a data type defined as
\linebreak \passthrough{\lstinline!data Nat = Zero | Suc Nat!}, the
equivalent extensible type in Haskell would be:

\begin{lstlisting}[language=Haskell]
data NatF f = ZeroF | SucF (f (NatF f))
\end{lstlisting}

We can use \passthrough{\lstinline!NatF Term!} to represent a natural
number where part of the definition is substituted by a variable. To
make the complete number replaceable with a variable, we enclose the
type inside an additional \passthrough{\lstinline!Term!}:

\begin{lstlisting}[language=Haskell]
type NatTerm = Term (NatF Term)
\end{lstlisting}

Examples of values of type \passthrough{\lstinline!NatTerm!} include
\passthrough{\lstinline!Var "y"!}, representing a natural number denoted
by the variable \passthrough{\lstinline!"y"!},
\passthrough{\lstinline!Compound (SucF (Var "x"))!}, representing the
successor of \passthrough{\lstinline!"x"!}, and
\passthrough{\lstinline!Compound (SucF (Compound ZeroF))!}, representing
\(1\). In Prolog, such terms could be encoded as
\passthrough{\lstinline!Y!}, \passthrough{\lstinline!suc(X)!}, and
\passthrough{\lstinline!suc(zero)!}, respectively.

\hypertarget{term-unification}{%
\subsection{Term Unification}\label{term-unification}}

We have designed a domain-specific language (DSL) to write predicates on
types for which we can perform term unification and variable
substitution. To unify two terms, we provide
\passthrough{\lstinline!(===) :: Term a -> Term a -> Goal!}.\footnote{The
  function is not fully polymorphic; we detail constraints applicable to
  the parameter \passthrough{\lstinline!a!} in Section
  \ref{implementation}.}

\hypertarget{example-2}{%
\paragraph{Example}\label{example-2}}

We can define the successor predicate as:\footnote{Compare with the
  Prolog program \passthrough{\lstinline!is\_suc(X,Y) :- suc(X) = Y.!}}

\begin{lstlisting}[language=Haskell]
isSuc :: NatTerm -> NatTerm -> Goal
isSuc x y = Compound (SucF x) === y
\end{lstlisting}

We now query this predicate in a session to check if it holds for two
\emph{ground} values:\footnote{We use \emph{ground} to refer to
  \emph{terms} that do not contain variables. The word ground also has
  meaning when discussing data types and generic programming, but we use
  the word exclusively with the former meaning.}

\begin{lstlisting}
> repl $ isSuc (Compound (SucF (Compound ZeroF)))
               (Compound (SucF (Compound (SucF (Compound ZeroF)))))
true.
> repl $ isSuc (Compound (SucF (Compound ZeroF)))
               (Compound (SucF (Compound ZeroF)))
false.
\end{lstlisting}

\noindent The real power of our approach is that we can now use
variables to provide one value and ``obtain'' the other, or rather, the
\emph{answer substitutions}:

\begin{lstlisting}
> repl $ isSuc (Compound (SucF (Compound ZeroF))) (Var "x")
x = Compound (SucF (Compound (SucF (Compound ZeroF)))).
> repl $ isSuc (Var "x") (Compound (SucF (Compound ZeroF)))
x = Compound ZeroF.
\end{lstlisting}

\hypertarget{existential-quantification}{%
\subsection{Existential
Quantification}\label{existential-quantification}}

The scope of variables as presented so far is, by default, global. A
variable with a fixed name being used inside a function will be
considered to be the same as a variable with the same name (and type)
used elsewhere in the same query.

To introduce free variables in the \emph{body} of predicates, akin to
introducing free variables in the antecedent in predicate definitions in
logic programming languages, we define the function
\linebreak \passthrough{\lstinline!exist :: (Term a -> Goal) -> Goal!}.
When the function \passthrough{\lstinline!exists!} is applied to an
argument predicate, it ensures that the variable provided to the given
predicate is free. To avoid name clashes, we recommend that users always
introduce variables with \passthrough{\lstinline!exists!}.

\hypertarget{example-3}{%
\paragraph{Example}\label{example-3}}

Using all the definitions provided so far, we can implement the
predicate \passthrough{\lstinline!leq!} (i.e., less than or equal to) to
compare two natural numbers, \passthrough{\lstinline!x!} and
\passthrough{\lstinline!y!}. The first rule of the comparison is that,
if the first number \passthrough{\lstinline!x!} is zero, then
\passthrough{\lstinline!leq x y!} must necessarily hold as there is no
smaller number, that is, \passthrough{\lstinline!x === Compound ZeroF!}.
The second rule is that, if both elements are successors of other
elements, respectively \passthrough{\lstinline!x'!} and
\passthrough{\lstinline!y'!}, then the goal holds if it holds for
\passthrough{\lstinline!x'!} and \passthrough{\lstinline!y'!}. Combining
both rules we obtain:

\begin{lstlisting}[language=Haskell]
leq :: NatTerm -> NatTerm -> Goal
leq x y = x === Compound ZeroF
       @| ( exists $ \x' -> exists $ \y' ->
            x === Compound (SucF x') @@ y === Compound (SucF y') @@ leq x' y' )
\end{lstlisting}

Introducing free variables at top of a definition helps group rules and
aids readability:

\begin{lstlisting}[language=Haskell]
leq :: NatTerm -> NatTerm -> Goal
leq x y = exists $ \x' -> exists $ \y' ->
     x === Compound ZeroF
  @| x === Compound (SucF x') @@ y === Compound (SucF y') @@ leq x' y'
\end{lstlisting}

As illustrated above, predicates can be \emph{recursive}.

\hypertarget{notation}{%
\subsection{Notation}\label{notation}}

The ideas of our proposal are applicable to other languages, but
Haskell's ability to overload notation and define operators can make
logic programming more convenient. Specifically, Haskell allows us to
define new operators and adjust their associativities and priorities,
making \passthrough{\lstinline!(@@)!} bind more strongly than
\passthrough{\lstinline!(@|)!}, and \passthrough{\lstinline!(===)!} bind
more strongly than either of them. We define synonyms
\passthrough{\lstinline!C!} and \passthrough{\lstinline!V!} for,
respectively, \passthrough{\lstinline!Compound!} and
\passthrough{\lstinline!Var!}, and pattern synonyms
\passthrough{\lstinline!Zero = C ZeroF!} and
\passthrough{\lstinline!Suc x y = C (SucF x y)!}. Using these
facilities, \passthrough{\lstinline!leq!} can now be defined more
succinctly as follows:

\begin{lstlisting}[language=Haskell]
leq :: NatTerm -> NatTerm -> Goal
leq x y = exists $ \x' -> exists $ \y' ->
     x === Zero
  @| x === Suc x' @@ y === Suc y' @@ leq x' y'
\end{lstlisting}

To help understand the technical details of our approach, we refrain
from relying too heavily on syntactic sugar during this exposition. Our
implementation provides aids to make using logic programming more
convenient, which we discuss in Section \ref{implementation}.

\hypertarget{polymorphism-and-higher-order}{%
\section{Polymorphism and Higher
Order}\label{polymorphism-and-higher-order}}

The ability to write predicates using the approach described so far
extends also to polymorphic types. Let us demonstrate with the type of
polymorphic lists, frequently used in logic and functional programs. The
standard list type definition,
\passthrough{\lstinline!data List a = Nil | Cons a (List a)!}, can be
extended with an extra type function as follows:

\begin{lstlisting}[language=Haskell]
data ListF f a = NilF | ConsF (f a) (f (ListF f a))
\end{lstlisting}

By applying the type function \passthrough{\lstinline!Term!} to
\passthrough{\lstinline!ListF!}, we can use logic variables in place of
elements of the list, or the tail of the list at any give point:

\begin{lstlisting}[language=Haskell]
type ListTerm a = Term (ListF Term a)
\end{lstlisting}

\hypertarget{example-4}{%
\paragraph{Example}\label{example-4}}

We can combine \passthrough{\lstinline!ListTerm!} with
\passthrough{\lstinline!NatF!} to represent lists of natural numbers:

\begin{lstlisting}
type NatListTerm = ListTerm (NatF Term)
\end{lstlisting}

The following, for now, rather verbose term encodes the list
\passthrough{\lstinline![0, 1, 2]!}:

\begin{lstlisting}
l1 :: NatListTerm
l1 = C $ ConsF (C ZeroF)
   $ C $ ConsF (C (SucF (C ZeroF)))
   $ C $ ConsF (C (SucF (C (SucF (C ZeroF))))) Nil
\end{lstlisting}

The following term encodes the list that starts with a 1, and whose tail
is represented by a variable \passthrough{\lstinline!"tl"!}:

\begin{lstlisting}
l2 :: NatListTerm
l2 = C (ConsF (C (Suc (C Zero))) (V "tl"))
\end{lstlisting}

\hypertarget{notation-1}{%
\paragraph{Notation}\label{notation-1}}

Like before, we introduce pattern synonyms to simplify writing terms of
type \linebreak \passthrough{\lstinline!ListTerm!}:
\passthrough{\lstinline!Cons x y = C (ConsF x y)!} and
\passthrough{\lstinline!Nil = C NilF!}. Using all pattern synonyms
defined so far, the list \passthrough{\lstinline!l1!} in the previous
example can be defined as:

\begin{lstlisting}
l1 :: NatListTerm
l1 = Cons Zero $ Cons (Suc Zero) $ Cons (Suc (Suc Zero)) Nil
\end{lstlisting}

Our implementation allows us to write expressions like
\passthrough{\lstinline![0, "x"]!} to mean a list with a first element
\(0\) and a second element being the variable
\passthrough{\lstinline!"x"!}. We delay notation aids to Section
\ref{implementation}, to help the reader gain intuition about how our
approach works.

\hypertarget{polymorphism-and-type-safety}{%
\subsection{Polymorphism and Type
Safety}\label{polymorphism-and-type-safety}}

Using extensible types does not prevent the host programming language
from performing type checking, including for types whose non-extensible
variants were polymorphic. If we try to use a term with the wrong type
in the definition of a predicate or in a unification constraint, the
type checker detects it just like it would any other type error.

\hypertarget{example-5}{%
\paragraph{Example}\label{example-5}}

Take the following predicate that checks if an element is the head of a
list:

\begin{lstlisting}[language=Haskell]
isHead :: ListTerm a -> Term a -> Goal
isHead x y = exists $ \tl -> x === Cons y tl
\end{lstlisting}

If we call \passthrough{\lstinline!isHead!} with a second argument of
the wrong type, the Haskell compiler's type checker warn:

\begin{lstlisting}[language=bash]
> repl $ isHead l1 ((Var "x") :: Term Bool)
<interactive>:9:19: error:
      Couldn't match type 'Bool' with 'NatF Term'
      Expected type: Term (NatF Term)
        Actual type: Term Bool
\end{lstlisting}

This level of safety is especially important when using free variables.
A host language with strong, static types can ensure that we are using
variables in type-consistent ways.

\hypertarget{example-6}{%
\paragraph{Example}\label{example-6}}

We can check if a value is in a list by checking against the head or
recursing into the tail. There is no rule for the empty list, which
makes the goal fail in that case:

\begin{lstlisting}[language=sh]
member :: Term a -> ListTerm a -> Goal
member x xs =
     ( exists $ \tl -> xs === Cons x tl )
  @| ( exists $ \hd -> exists $ \tl -> xs === Cons hd tl @@ member x tl )
\end{lstlisting}

The variable \passthrough{\lstinline!hd!} in the second rule has type
\passthrough{\lstinline!Term a!}, and \passthrough{\lstinline!tl!} has
type \passthrough{\lstinline!List a!}. The compiler can infer this
because both are arguments to \passthrough{\lstinline!Cons!} and the
resulting term unifies with \passthrough{\lstinline!xs!}, whose type is
known. If, for example, we introduce a condition
\passthrough{\lstinline!xs === hd!}, or
\passthrough{\lstinline!member tl tl!}, the compiler will detect that we
are using variables in inconsistent ways.

Let us further demonstrate the type safety features of our approach with
a predicate that checks if a list is sorted. The predicate holds
trivially for lists of zero or one elements; if there are more, we check
the first two and recurse into the tail of the list.

\begin{lstlisting}[language=Haskell]
sorted v =
     v === Nil
  @| ( exists $ \e1 -> v === Cons e1 Nil)
  @| ( exists $ \e1 -> exists $ \e2 -> exists $ \ts ->
            v === Cons e1 (Cons e2 ts)
         @@ leq e1 e2
         @@ sorted (Cons e2 ts) )
\end{lstlisting}

In this case, the compiler infers that \passthrough{\lstinline!v!} has
type \passthrough{\lstinline!NatList!}, which is defined as
\linebreak \passthrough{\lstinline!ListTerm (NatF Term)!} and expands to
\passthrough{\lstinline!Term (ListF Term (NatF Term))!}.

\hypertarget{higher-order-logic-programming}{%
\subsection{Higher-order Logic
Programming}\label{higher-order-logic-programming}}

Haskell's support for first-class functions immediately empowers our
approach with higher-order, allowing us to pass predicates as arguments
to other predicates.

\hypertarget{example-7}{%
\paragraph{Example}\label{example-7}}

Let us illustrate with a generalized version of
\passthrough{\lstinline!sorted!} that takes a comparison
\emph{predicate} as argument:

\begin{lstlisting}[language=sh]
sortedWith :: (Term a -> Term a -> Goal) -> ListTerm a -> Goal
sortedWith compare v =
     v === Nil
  @| ( exists $ \x -> v === Cons x Nil )
  @| ( exists $ \x1 -> exists $ \x2 -> exists $ \xs ->
            v === Cons x1 (Cons x2 xs)
         @@ compare x1 x2
         @@ sortedWith compare (Cons x2 xs) )
\end{lstlisting}

Because Haskell is strongly and statically typed, it provides a level of
safety that surpasses what most implementations of Prolog
offer,\footnote{Some Prolog implementations, like Ciao Prolog, support
  static analysis via compile-time assertions.} since they cannot assure
that types match without added, hand-coded runtime checks. In general,
calling a Prolog predicate with arguments of the wrong types may return
an incorrect result, making this kind of type error hard to identify.
Providing arguments of the wrong type may return
\passthrough{\lstinline!false!}, just as if the predicate did not hold
for those inputs (because it does not!), but may also incorrectly return
\passthrough{\lstinline!true!} (e.g.,
\passthrough{\lstinline!append([],1,1)!} is
\passthrough{\lstinline!true!} even though \passthrough{\lstinline!1!}
is not a list).

We can use the same approach to generalize functions and turn them into
predicates, such as the standard function \passthrough{\lstinline!map!}
that applies a transformation to each element in a list, as illustrated
by the following predicate \passthrough{\lstinline!mapP!}:

\begin{lstlisting}[language=Haskell]
mapP :: (Term a -> Term b -> Goal) -> ListTerm a -> ListTerm b -> Goal
mapP f l1 l2 =
     l1 === Nil @@ l2 === Nil
  @| ( exists $ \l10 -> exists $ \l1s -> exists $ \l20 -> exists $ \l2s ->
            l1 === Cons l10 l1s
         @@ l2 === Cons l20 l2s
         @@ f l10 l20 @@ mapP f l1s l2s )
\end{lstlisting}

Given a predicate \passthrough{\lstinline!isSuc!}, which pairs each
number with its successor, we can use it to add \(1\) to every element
of a list using \passthrough{\lstinline!mapP!} as follows:

\begin{lstlisting}[language=Haskell]
listPlusOne :: ListTerm (NatF Term) -> ListTerm (NatF Term) -> Goal
listPlusOne = mapP isSuc
\end{lstlisting}

Haskell's type checker prevents runtime errors by ensuring, at compile
time, that \passthrough{\lstinline!isSuc!} has type
\passthrough{\lstinline!Term (NatF Term) -> Term (NatF Term) -> Goal!}.

\hypertarget{cuts}{%
\section{Cuts}\label{cuts}}

In logic programming, \emph{cuts} limit the use of backtracking to
search for alternative solutions. Consider the following (incorrect)
implementation of the remainder algorithm:

\begin{lstlisting}
remainder :: NatTerm -> NatTerm -> NatTerm -> Goal
remainder n q r = lt n q @@ n === r
               @| exists $ \diff -> plus q diff n @@ remainder diff q r
\end{lstlisting}

This definition does not work if \passthrough{\lstinline!q!} is zero,
since the first rule fails, and the execution of the second rule leads
the program into an infinite loop. Unfortunately, adding a protection
rule like \passthrough{\lstinline!q === Zero @@ fail!} does not help,
since, if \passthrough{\lstinline!q!} is zero, that rule fails and the
evaluation backtracks, eventually falling into the last rule again.

To prevent such cases, we introduce the functions
\passthrough{\lstinline!scope!} and \passthrough{\lstinline"(@!)"},
which help control backtracking. Inspired by the notion of cuts in
Prolog, we refer to \passthrough{\lstinline"(@!)"} as our own cut
operator. We re-write \passthrough{\lstinline!remainder!} using these
two functions as follows:

\begin{lstlisting}
remainder :: NatTerm -> NatTerm -> NatTerm -> Goal
remainder n q r =
  scope $ q === Zero @! fail
       @| lt n q @@ n === r
       @| (exists $ \diff -> plus q diff n @@ remainder diff q r)
\end{lstlisting}

Callers to \passthrough{\lstinline!remainder!} are unaware that the
predicate is implemented using cuts, provided that uses of
\passthrough{\lstinline"(@!)"} are delimited by
\passthrough{\lstinline!scope!}, limiting how far the cut applies.
Without \passthrough{\lstinline!scope!} in the definition of
\passthrough{\lstinline!remainder!}, predicates using
\passthrough{\lstinline!remainder!} in a rule would see alternative
(i.e., \passthrough{\lstinline!(@|)!}) rules being skipped over if
\passthrough{\lstinline!remainder!} fails due to
\passthrough{\lstinline!q!} being \passthrough{\lstinline!Zero!}.

\hypertarget{negation-as-failure}{%
\paragraph{Negation as Failure}\label{negation-as-failure}}

Cuts can be used to implement a form of negation, with:

\begin{lstlisting}[language=sh]
neg :: Goal -> Goal
neg p = scope $ p @! fail @| succeed
\end{lstlisting}

\hypertarget{example-8}{%
\paragraph{Example}\label{example-8}}

It is frequently useful to state that two terms cannot unify, for which
we define an operator \passthrough{\lstinline!(=/=)!} as:

\begin{lstlisting}[language=sh]
(=/=) :: Term a -> Term a -> Goal
(=/=) x y = neg (x === y)
\end{lstlisting}

Similarly, we can implement a predicate that holds only if a given term
is \emph{not} a member of a given list:

\begin{lstlisting}[language=sh]
notMember :: Term a -> List a -> Goal
notMember x xs = neg (member x xs)
\end{lstlisting}

This kind of negation, called \emph{negation as failure}, is a weak form
of negation. If either of the arguments of
\passthrough{\lstinline!(=/=)!} is still a variable at the time when the
unification algorithm tries to evaluate whether the goal holds, the
unification \passthrough{\lstinline!x === y!} will hold, making its
negation fail. This limits the usefulness of this form of negation. For
example, one cannot use \passthrough{\lstinline!notMember!} as defined
above to find possible values of a variable that are not members of a
list. Instead, it is necessary to \emph{ground} the term first. To this
end, we provide the predicate \passthrough{\lstinline!isGround!} which,
for any type for which ground terms have a finite representation, holds
only if the given argument contains no variables.

\hypertarget{implementation}{%
\section{Implementation}\label{implementation}}

We have implemented the ideas in this paper in Haskell\footnote{Our
  implementation has been made publicly available at
  https://github.com/ivanperez-keera/telos}, including types
representing \passthrough{\lstinline!Term!}s and
\passthrough{\lstinline!Goal!}s; goal building functions and
combinators; classes that define the operations that types must support
for unification to be used on them; a unification algorithm; and
execution functions to evaluate goals. Overall, our implementation only
needs 350 lines of code, without considering spaces or comments. We have
implemented this solution with the aim of demonstrating the capabilities
explained, and explore design decisions, syntax and embeddings.
Evaluating the performance of the solution and comparing it with
existing logic programming implementations is out of the scope of this
paper and considered future work.

\hypertarget{high-level-description}{%
\paragraph{High-level Description}\label{high-level-description}}

Our implementation defines two key types: a polymorphic type
\passthrough{\lstinline!Term!}, described in Section
\ref{logic-programming-with-extensible-types}, and a type
\passthrough{\lstinline!Goal!}, which represents a goal. To perform
unification for terms of arbitrary types, we require that three
operations be supported on \passthrough{\lstinline!Term!}s: the ability
to unify two terms of a specific type, the ability to check if a
variable is used in a term, and the ability to substitute a variable by
a term inside another term. We capture these operations in Haskell via
type classes, the key one being the type class
\passthrough{\lstinline!Logic!}, which represents types for which the
aforementioned operations are defined.

To evaluate goals, we provide several functions, including:
\passthrough{\lstinline!repl :: Goal -> IO ()!}, which prints solutions
one by one, letting users control the production of solutions with the
keyboard, and
\passthrough{\lstinline!findAll :: Logic a => Term a -> Goal -> [Term a]!},
which provides all values for an argument variable
\passthrough{\lstinline!Term!} under which a
\passthrough{\lstinline!Goal!} may hold.

Our implementation represents goals using a tree-like structure with
unification constraints in the nodes. An internal function
\passthrough{\lstinline!solve!} traverses the tree, accumulating
unification constraints, substituting variables with their expected
values in other terms, and discarding \linebreak branches that cannot be
unified. The function \passthrough{\lstinline!solve!} produces all
possible solutions, each of which assigns values to variable terms. Our
function \passthrough{\lstinline!repl!} leverages Haskell's inherent
laziness to produce and print solutions one by one.

\hypertarget{notation-and-usability}{%
\paragraph{Notation and Usability}\label{notation-and-usability}}

To make our DSL convenient, we have implemented facilities to make the
notation succinct and familiar, and reduce how much code users must
write. Our implementation uses generic programming
\citep{2010:magalhaes:generics} to generate instances of the classes
that our unification function requires to operate on algebraic data
types. For example, for the type \passthrough{\lstinline!NatF!}, users
need to declare several type class instances, but do not need to
implement them manually. Our solution also facilitates writing terms of
some common types. For example, Peano natural numbers are printed as
\passthrough{\lstinline!1!}, \passthrough{\lstinline!2!}, etc. when they
are ground terms, and a number plus a variable otherwise (e.g.,
\passthrough{\lstinline!1 + x!}, \passthrough{\lstinline!55 + z!}). We
implement similar aids for \passthrough{\lstinline!ListF!}, so that
values can be shown and given in a familiar notation (e.g.,
\passthrough{\lstinline!1 : x : 5 : xs!}). Where a
\passthrough{\lstinline!Term!} is expected, a literal string is
interpreted as variable term. Overall, this renders very concise
expressions:

\begin{lstlisting}[language=Haskell]
> repl (plus 1 "x" 5)
x = 4.
> repl (isTail [1, 2, 3] [2, 3])
true.
\end{lstlisting}

The same mechanisms for syntax overloading may not be available in other
languages that otherwise support higher-kinded polymorphism or
dynamically replacing a value by a sum type, meaning that the notation
in such languages could be more cumbersome.

\hypertarget{related-work}{%
\section{Related Work}\label{related-work}}

\hypertarget{functional-logic-programming-languages}{%
\paragraph{Functional-Logic Programming
Languages}\label{functional-logic-programming-languages}}

The creation of languages that integrate logic and functional
programming using theoretical frameworks and efficient implementations
has been subject to prior study \citep{1986:flp, 1986:lindstrom}.
Languages in this category include Babel \citep{1988:babel}, K-LEAF
\citep{1991:kleaf}, ALF \citep{1991:alf}, Curry
\citep{1995:moreno:curry}, and Escher \citep{1999:lloyd:escher}, which
support a functional style, and Gödel \citep{1994:lloyd:godel}, Mercury
\citep{1996:mercury}, and \(\lambda\)Prolog
\citep{1986:nadathur:higherorderprolog}, which embrace a logic
programming style. Curry, in particular, is strongly inspired by Haskell
but incorporates logical variables and uses narrowing as operational
semantics to compute the value of expressions with free variables
\citep{2013:hanus:flp_curry}. Instead of creating a new language, our
work shows how an existing functional language can be empowered with
logic programming capabilities, without additional compiler extensions
and without calling an external logic programming engine.

\hypertarget{functions-in-logic-programming-languages}{%
\paragraph{``Functions'' in Logic Programming
Languages}\label{functions-in-logic-programming-languages}}

Prolog includes limited higher-order capabilities like
\passthrough{\lstinline!call/N!} and \passthrough{\lstinline!apply/3!}
\citep{1996:naish:higherorderprolog}. The Prolog implementation Ciao
\citep{2012:ciaophilosophy} allows, via its metaprogramming libraries,
using predicates in a functional style, treating the last argument as
the result of the function. In our case, functions and higher-order come
built-in with the host language and are immediately exploitable by
programmers. Furthermore, because we rely on a statically typed
language, our approach provides a level of static safety.

\hypertarget{typed-logic-programming-languages}{%
\paragraph{Typed Logic Programming
Languages}\label{typed-logic-programming-languages}}

Prolog has previously been extended with static types
\citep{1984:mycroft, 2008:typedprolog, 2022:prologtypeinference}, but
these extensions are not integrated in most widely used Prolog systems.
The implementation Ciao Prolog \citep{2012:ciaophilosophy} provides a
mechanism of assertions based on \emph{regular types} that are checked
statically. In contrast, our proposal leverages the type system of the
host programming language.

\hypertarget{logic-programming-embeddings}{%
\paragraph{Logic Programming
Embeddings}\label{logic-programming-embeddings}}

Prior attempts at embedding logic programming in functional languages by
\citet{1999:spivey:prologinhaskell}, \citet{2001:koen:prologinhaskell},
\citet{2012:solanki:prologinhaskell} and \citet{1991:elliott:logicprog}
require adapting the types by hand to use them in logical predicates.
Work by \citet{2016:kosarev:embedding_relational_in_ocaml} to embed
relations in O'Caml require introducing \emph{projections} and
\emph{injections} to move between terms and functional values. In
contrast, our approach is applicable to arbitrary algebraic data types,
with generic programming aids to facilitate operating with them,
simplifying the process. The application of a systematic extension
pattern, rather than hand-coded extensions, leads to regular and
predictable ways to add variables to algebraic data types.

In terms of the implementation of goal evaluation, prior embeddings use
an interpreter based on an evaluation monad
\citep{1999:spivey:prologinhaskell, 2001:koen:prologinhaskell, 2012:solanki:prologinhaskell, 2009:schrijvers:monadiccp},
make goals data \emph{streams} and explicitly introduce a backtracking
lazy stream monad in an otherwise strict setting
\citep{2016:kosarev:embedding_relational_in_ocaml}, or use continuation
passing style and exceptions to evaluate logic programs more efficiently
\citep{1991:elliott:logicprog}. In contrast, we use an internal
representation of goals as a tree of unification constraints,
parameterized by a counter used to generate free variables. Our encoding
provides a more fine-grained control of the counter than what we could
obtain by using a (lazy) state monad, and introduces a level of laziness
that is crucial to generate solutions efficiently and implement the
REPL. Another difference in implementation between miniKanren
\citep{2016:kosarev:embedding_relational_in_ocaml} and our work is that
the former uses a type-unsound internal representation encapsulated
behind a type-safe API. Our approach implements unification without
sacrificing type safety, and term unification can only be applied to
terms of the same type.

These embeddings also differ from our work in terms of expressiveness:
for example, we only support equality constraints, and implement a
limited form of disequality using a (weak) form of negation via cuts,
whereas miniKanren and the work of \citet{2009:schrijvers:monadiccp}
support both equality and disequality.

A more recent publication presents typedKanren
\citep{2024:kudasov:typedkanren}, an embedding of a relational language
based on miniKanren inside Haskell. Like our proposal, typedKanren
enables writing strongly typed predicates in Haskell, and uses a type
for term that is similar to ours (except for added strictness and the
use of \passthrough{\lstinline!Int!} instead of
\passthrough{\lstinline!String!} to represent variables). A key
difference is that typedKanren requires
\passthrough{\lstinline!Logical!} variants of types to be written for
standard types (e.g., binary tries), whereas our representations are
based on extensible types and are therefore more versatile. Like
miniKanren, typedKanren supports disequalities, while we do not support
them in the general case. Goals are monads in typedKanren, which results
in a different style when specifying predicates, while goals in our
language are not monads, and we instead provide additional constructs
and syntactic aids to make the notation more convenient and closer to
the syntax of Prolog.

\hypertarget{higher-kinded-type-parametrizations}{%
\paragraph{Higher-kinded Type
Parametrizations}\label{higher-kinded-type-parametrizations}}

Our solution uses extensible types to parameterize types with type
functions. \citet{2016:najd:trees} propose an alternative technique that
uses \emph{type families} and adds a parameter
\passthrough{\lstinline!f!} to every branch of an algebraic data type's
definition, as opposed to every element inside the type. The work of
Najd et al.~also allows introducing \passthrough{\lstinline!Term!}-like
wrappers around all elements of the abstract data type. However, we find
extensible types straightforward due to the minimal work required to
replace \emph{any} part of a type with a variable by applying
\passthrough{\lstinline!Term!} to the extensible type.

The library \passthrough{\lstinline!barbies!} \citep{barbies} implements
generic mechanisms to work with types parameterized by a functor similar
to extensible types, and Barbies-th \citep{barbies-th} or Higgledy
\citep{higgledy} could help generate higher-kinded types in Haskell. We
have yet to investigate how to take advantage of such approaches.

\hypertarget{future-work}{%
\section{Future Work}\label{future-work}}

This paper has shown how to embed logic programming in a statically
typed functional programming language. To that end, we used extensible
types to replace any portion of an algebraic datatype with variables,
and provided a mechanism to express unification constraints between
values with variables. We further extended the language with boolean
connectives, and ways to introduce free variables. We showed by example
that we can leverage the host language's type inference and higher-order
to make code reusable without sacrificing type safety. We closed our
discussion with an overview of our implementation, and an evaluation of
the differences with other approaches.

The approach proposed in this paper could be used to represent other
kinds of constraints, allowing us to implement constraint-logic
programming in Haskell. Introducing disequality constraints in our
language would overcome the limitations of our current implementation of
disequality using ``negation as failure''.

We are currently exploring how the language provided in this paper could
be used to generate values that meet constraints in property-based
testing, rather than first generating values and then filtering based on
constraints (thus discarding many values).

Our experiments indicate that the semantics of our goal evaluation
functions coincides with that of Prolog. We would like to carry out a
more detailed and formal evaluation to compare our inference engine with
Prolog's. Additionally, we have not discussed benchmarking our
unification algorithm against existing implementations, as it is a topic
that deserves careful and detailed evaluation, and we consider future
work.

In future work, we would like to make our DSL more user-friendly and the
notation closer to that of Prolog or other logic programming languages,
without sacrificing the embedding in a host language. Prolog predicates
are normally defined by multiple rules, where the head of the rule can
unify with the arguments. In contrast, in our host language, pattern
matching cannot be used to perform unification, so we need to use
\passthrough{\lstinline!(===)!} in definitions to split the inputs into
their component parts. We consider leveraging compiler extensions to
mimic Prolog's unification of clause heads as future work.

In our proposal, users have to define an extensible type counterpart of
the types they want to work with, wrap values into terms, and transform
ground terms back into values of the original, non-extensible type.
Although our notation aids partly simplify that process, we plan to
investigate how template meta-programming could help generate the
extensible type associated to a given algebraic datatype automatically.

Prolog allows using underscores in place of terms, to indicate that the
value in place should unify, but not capture it, equivalent to
introducing free variables where underscores are used. Our
implementation supports wildcard patterns, or underscores, on the
left-hand side of a predicate definition. In the future, we plan to
investigate how to support wildcard patterns also in the right-hand side
of predicate definitions.

Finally, we plan to evaluate how to remove the need to manually add
\passthrough{\lstinline!scope!} to limit the effect of cuts in our
language.

\clearpage

\bibliographystyle{eptcs}
\interlinepenalty=10000
\bibliography{paper}

\end{document}